# Equivalence testing in pesticide risk assessment - Evaluation and practical guidance for design, analysis and interpretation

Dimitry Wintermantel[1*], Julia Osterman[2,3,4], Magdalena M. Mair[5], Florian Hartig[6]

[1] Chair for Nature Protection and Landscape Ecology, University of Freiburg

[2] Department of Biological & Environmental Sciences, University of Gothenburg, Sweden

[3] Centre for Future Chemical Risk Assessment and Management (FRAM), University of Gothenburg, Sweden

[4] Gothenburg Global Biodiversity Centre, Gothenburg, Sweden

[5] Statistical Ecotoxicology, Bayreuth Center of Ecology and Environmental Research (BayCEER), University of Bayreuth

[6] Theoretical Ecology, Faculty of Biology and Preclinical Medicine, University of Regensburg

***Corresponding author**: dimitry.wintermantel@nature.uni-freiburg.de

**Keywords**: Equivalence test, environmental risk assessment, plant protection product, non-inferiority test, point-null test, *Apis mellifera*, pollinator

## Abstract

Harmful pesticide effects exceeding specific protection goals (SPG) may go undetected in underpowered experimental designs. Regulatory honeybee field studies have consistently failed to reach the statistical power required under European Food Safety Authority (EFSA) guidance, which may have caused approval of high-risk substances. Therefore, EFSA advised a shift from testing the null hypothesis of 'no effect' to equivalence testing. Under this approach, a pesticide is classified as 'low risk' if the null hypothesis that its effect exceeds the SPG can be rejected. For honeybees, the recommended SPG is a colony size reduction below 10%. Critics have argued that this framework requires excessive site replication to demonstrate pesticide safety and proposed an alternative equivalence test defining treatment effects relative to the lower bound of the 90%-control-group confidence interval. Using simulations mimicking a regulatory honeybee field study, we show that although the two equivalence tests share the same trade-off between false 'low-risk' and false 'high-risk' classifications, only EFSA's original recommendation reliably identifies pesticides with effects > SPG at $\alpha$=0.2. Our results show that increasing site replication beyond the current practice is unavoidable for a reliable regulatory assessment. However, for pesticides with effect sizes ≤5%, site requirements remain lower than those implied by the power requirement of the former EFSA guidance. Moreover, covariate adjustment—through a model term or balanced colony allocation using anticlustering randomisation—can reduce site requirements without losing power and thus save costs. Finally, we provide guidance and R functions for anticlustering randomisation and equivalence testing for pesticide risk assessment.

# 1 Introduction

Pesticides harm non-target organisms across different taxa worldwide (Wan et al., 2025) even though in most countries pesticides undergo ecological risk assessments before their approval (Krahner et al., 2022). These risk assessments often follow a tiered approach, whereby pesticides are first screened under controlled laboratory conditions and, if potential risks are indicated, tested under increasingly realistic conditions in higher tiers. Indications of high risk at lower tiers trigger testing at higher tiers so that ultimately approval based on ecological risk is only denied at the highest tier (Krahner et al., 2022; Sánchez-Bayo and Tennekes, 2015). In higher-tier studies, a single test group exposed to a realistic pesticide dose/concentration is compared with a control group. Traditionally, such studies have relied on significance testing against the null hypothesis of 'no effect' (hereafter '*difference tests*'), where a failure to detect significant negative impacts is often taken as evidence for the tested substance being safe. However, in such an approach, even large detrimental effects may go unnoticed in studies that lack statistical power due to high variance or low replication, even though the weight-of-evidence framework used to integrate across studies considers effect sizes (EFSA Scientific Committee et al., 2017). The framework formally considers study reliability, but the qualitative nature of reliability scoring means that imprecision and non-significant results in individual studies may influence conclusions. Underpowered studies may be part of the reason why pesticides that passed the bar for approval in the European Union were, in many cases, later found to be harmful and consequently restricted or banned (Storck et al., 2017).

To avoid this problem, regulatory protocols usually include sample size or power requirements. However, this does not necessarily ensure adequate protection against this power problem, as illustrated by official guidance on honeybee testing. Honeybees are the standard model species for pollinator risk assessments, with field studies representing the highest tier in both the European Union and the United States (EFSA 2023; US EPA 2014). While the internationally recognised guideline EPPO 1/170, published in 2010, requires only four colonies per site and no site replication, the European Food Safety Authority (EFSA) introduced in 2013 in their guidance document for bees a power requirement specifying that studies should be designed to detect a 7% reduction in colony size (measured as the number of adult bees per colony at the end of the test) with 80% power (Table 1). However, this guidance lacked support from EU Member States, preventing it from becoming the standard risk assessment procedure (EFSA 2023). Nevertheless, the guidance document was largely followed in the 2018 reassessment of three neonicotinoid insecticides (EFSA 2018a, EFSA 2018b, EFSA 2018c), except that none of the assessed field studies met the power requirements due to insufficient replication (EFSA 2021; Woodcock et al., 2016). In fact, it was estimated that none of 63 semi-field and field studies on neonicotinoid insecticides had sufficient power to detect a 7% reduction, and fewer than 27% of studies had the power to detect a large reduction of >35% of worker bees (Franklin and Raine, 2019). This lack of power in higher-tier studies may help explain why multiple pesticides harmful to pollinators have passed regulatory risk assessments in the EU and beyond, with some later banned after impacts became evident post-approval (Nicholson et al., 2024; Pisa et al., 2021; Rundlöf et al., 2015; Wintermantel et al., 2018).

To address these problems, in 2023, EFSA released a revision of its guidance on the risk assessment of plant protection products for bees. The revised guidance includes a conceptual shift from difference testing to equivalence testing (EFSA 2023). This shift effectively changes the direction of the statistical tests: Under the old paradigm, companies seeking approval for their pesticide had to show evidence that there was no negative effect. Under the new paradigm, they must provide evidence that negative effects are lower than a given threshold. More specifically, 'low risk' is concluded when a one-sided equivalence test (essentially a non-inferiority test) shows that the reduction in honeybee colony size between the pesticide and control group is significantly smaller than the 10% threshold, defined as the specific protection goal (SPG) in the revised guidance document at a significance level *alpha* ($\alpha$) of 0.2.

**Table 1. Summary of regulatory guidance documents (GD) / protocols with respect to requirements in the statistical analyses and sample size of field studies.**

| | **EPPO 1/170 (2010)** | **EFSA GD (2013)** | **EFSA GD (2023)** |
|---|---|---|---|
| **Specific protection goal (SPG)** | Not defined | Max 7% reduction in colony size | Max 10% reduction in colony size |
| **Low risk is concluded if** | No information | No significant effect (one-sided point difference test, $\alpha = 0.05$) | Significant one-sided equivalence test with threshold = 10% ($\alpha = 0.2$) |
| **Required statistical power** | None | 80% to achieve SPG | Implied |
| **Recommended sample size** | 1 field with min 4 colonies in each treatment | 14 fields with 7 colonies in each treatment | 10 fields with 6 colonies in each treatment (for effect size = 0%) |

In evaluating the practical implications of this new policy for the disapproval rate, risk assessors from both the public and private sectors (Hotopp et al., 2024) examined the approach by simulating field studies based on typical control honeybee colony data from Rolke et al. (2016) with a true effect size of 0%. They found that in most cases, an equivalence test was non-significant, which they interpreted, following EFSA's guidance, as a false 'high risk' conclusion. Due to the high rate of non-significant results, Hotopp *et al*. (2024) concluded that with the new EFSA guidelines, it is too difficult to meet the criteria necessary for regulatory approval in higher-tier field experiments. Therefore, they proposed an altered equivalence test where the baseline for determining colony size reduction is shifted from the control mean to the lower bound of a 90% confidence interval around the control mean (hereafter '*Hotopp test*' or '*Hotopp equivalence test'*), arguing that this would account for the natural variability of honeybee colonies. They showed that with this test, a larger rate of harmless substances would pass the regulatory assessment and thus argued for altering the original EFSA recommendation.

We are concerned with this recommendation for several reasons. First, the high rate of 'high risk' findings observed by Hotopp *et al.* is unsurprising, as they simulated data for underpowered studies with only half of the original sample size from Rolke *et al.* (2016) - itself a typical honeybee field study with just six sites per treatment, which was already below the ten sites that EFSA estimated would be needed to reliably demonstrate low risk when the true effect size is 0%. The high non-significance rates are therefore completely expected and could be resolved by increasing the sample size.

More importantly, however, in statistics there is rarely a free lunch – even without knowing their methods in detail, the fact that their altered test resulted in a higher rate of studies passing the test should make us wary about whether the new proposal just lowers the stringency of the test. Minimising false 'high risk' conclusions is important, of course, but this should not be achieved at the cost of increased rates of false trust; that is, mistakenly concluding that a pesticide is safe when it poses unacceptable risks. Avoiding false trust is particularly important in higher-tier studies because these are usually initiated only after lower tiers have indicated the potential for a high risk. Therefore, it is important to understand how the test proposed by Hotopp *et al.* differs from the EFSA equivalence test in the ability to distinguish between high-risk and low-risk pesticides and to assess what the implications of adopting this test in the regulatory risk assessment would be particularly with respect to the approval rates of pesticides with effect sizes that exceed the tolerable threshold (*i.e*. the SPG).

Irrespective of whether or not the altered test proposed by Hotopp *et al.* is valid, we agree with their concern that it is important to maximise power in honeybee field studies, as higher power improves the feasibility of approving low-risk pesticides and may thereby increase the proportion of approved pesticides that are truly low risk. One cost-effective way of increasing power may be to account for variables that introduce noise - either statistically or through optimised experimental design. Completely random allocation of colonies to treatments ensures that the treatments are uncorrelated to other variables measured before treatment application, particularly in infinitely large studies. However, in experiments with a limited sample size, imbalances in influential variables among treatments may occur. Such biases may lead to differences in measured endpoints between treatments that are not consequences of the treatments themselves and introduce variance to the test system, ultimately lowering statistical power. Understanding the power gains from minimising such imbalances or statistically adjusting for them can provide valuable guidance for both study design and data analysis.

Here, we compare the equivalence tests proposed by EFSA and Hotopp *et al.* in terms of their ability to distinguish between high-risk and low-risk pesticides and more specifically to protect against falsely classifying high-risk pesticides as 'low risk'. To understand the required level of replication under EFSA guidance and reach recommendations for approval for pesticides that cause little to no harm to honeybee colonies, we then determine the number of sites required to demonstrate that pesticides with true effect sizes of less than 10% in magnitude are 'low risk' under EFSA's current equivalence-based guidance, and compare these sample size requirements to those under EFSA's former difference-testing approach (with -7% threshold for the effect size) (EFSA 2013; EFSA 2023). Next, we examine how many fewer sites are required if initial colony size is adjusted for, either by including it as a fixed effect in the model structure during analysis or by using anticlustering randomisation to optimise colony allocation to sites and treatments at the start of experiments via a new R function, or both. Finally, we provide guidance on how to implement both anti-clustering randomisation and equivalence testing in R, as well as recommendations for interpreting equivalence test results in risk assessments.

# 2 Methods

## 2.1 Original dataset

As a base for our simulations, we used a field experiment by Rolke *et al*. (2016) that was submitted by the industry to regulatory authorities for the reapproval of a neonicotinoid insecticide for the EU market. The experiment investigated the effects of this pesticide on honeybee colony size/strength, measured as the number of (adult) bees (per colony) and consisted of a total of twelve sites (referred to as 'locations' in Rolke *et al.*), half of which were at or near (i.e. approximately 400 meters away from) neonicotinoid-treated oilseed rape fields or at or near oilseed rape fields that were not treated with a neonicotinoid (control sites). For our simulations, we used only the control data and, following Rolke *et al.*'s approach, we did not differentiate between sites located directly at oilseed rape fields and those 400 meters away, treating each site as an independent unit.

Rolke *et al.* assessed the number of bees per colony at four time points during the oilseed rape bloom but not before pesticide application, as it is recommended in official guidelines (EPPO 2010; EFSA 2013; EFSA 2023). Therefore, we used data from the first assessment time point during the oilseed rape bloom to determine the initial number of bees, and the second, third and fourth time points to simulate colony development over time.

## 2.2 Simulation procedure

To generate new datasets of desired size with similar properties as in Rolke *et al.*, we used a parametric bootstrap. The risk simulation procedure consisted of the following steps: i) derive distributional parameters from original control data, ii) simulate a dataset of desired size with colony identity, site identity and initial number of bees per colony, iii) optionally reallocate colonies using anticlustering randomization based on initial number of bees, iv) simulate data for post-application assessment time points, assign treatments (control and pesticide) to sites and apply a treatment effect of predetermined size to sites of the pesticide group by reducing the number of bees per colony by a percentage corresponding to the magnitude of the effect size, v) fit the generated data to a generalized mixed-effects model (GLMM) and iv) evaluate the model to classify pesticides either as 'low risk' or 'high risk' using one or more one-sided tests of the following: standard equivalence testing as suggested by EFSA, modified equivalence test proposed by Hotopp *et al.* and difference testing (Fig. 1).

The procedure was repeated 5,000 times for each considered effect size and each tested covariate adjustment strategy for initial number of bees to estimate the percentage of simulations in which the pesticide was classified as 'low risk'. Up to four adjustment strategies were tested: i) no adjustment: not accounting for initial number of bees, ii) design adjustment: balanced colony allocation across treatments and sites using anticlustering randomization based on initial numbers of bees during experimental setup, iii) statistical adjustment: including initial number of bees as covariate (*i.e*. continuous fixed effects variable) in the GLMM during analysis, iv) combined adjustment: balanced colony allocation across treatments and sites during experimental setup and inclusion of initial number of bees as covariate during analysis.

For verification of the risk simulation process, we slightly modified the process to illustrate that applying it to the original data yields very similar results to when simulating new data with the same number of sites and colonies per site (Fig. A1). Two sites were randomly assigned to the pesticide group and one to the control group. To introduce variability across iterations while preserving the underlying data structure, we permuted the site assignments of colonies while maintaining the treatment assignment for each site. After this permutation, the treatment effect was applied to the pesticide group. For comparability, the colony permutation was done both with the original data and the simulated data.

To reduce computational time, we parallelised the simulation runs (for details see Supplementary Material).

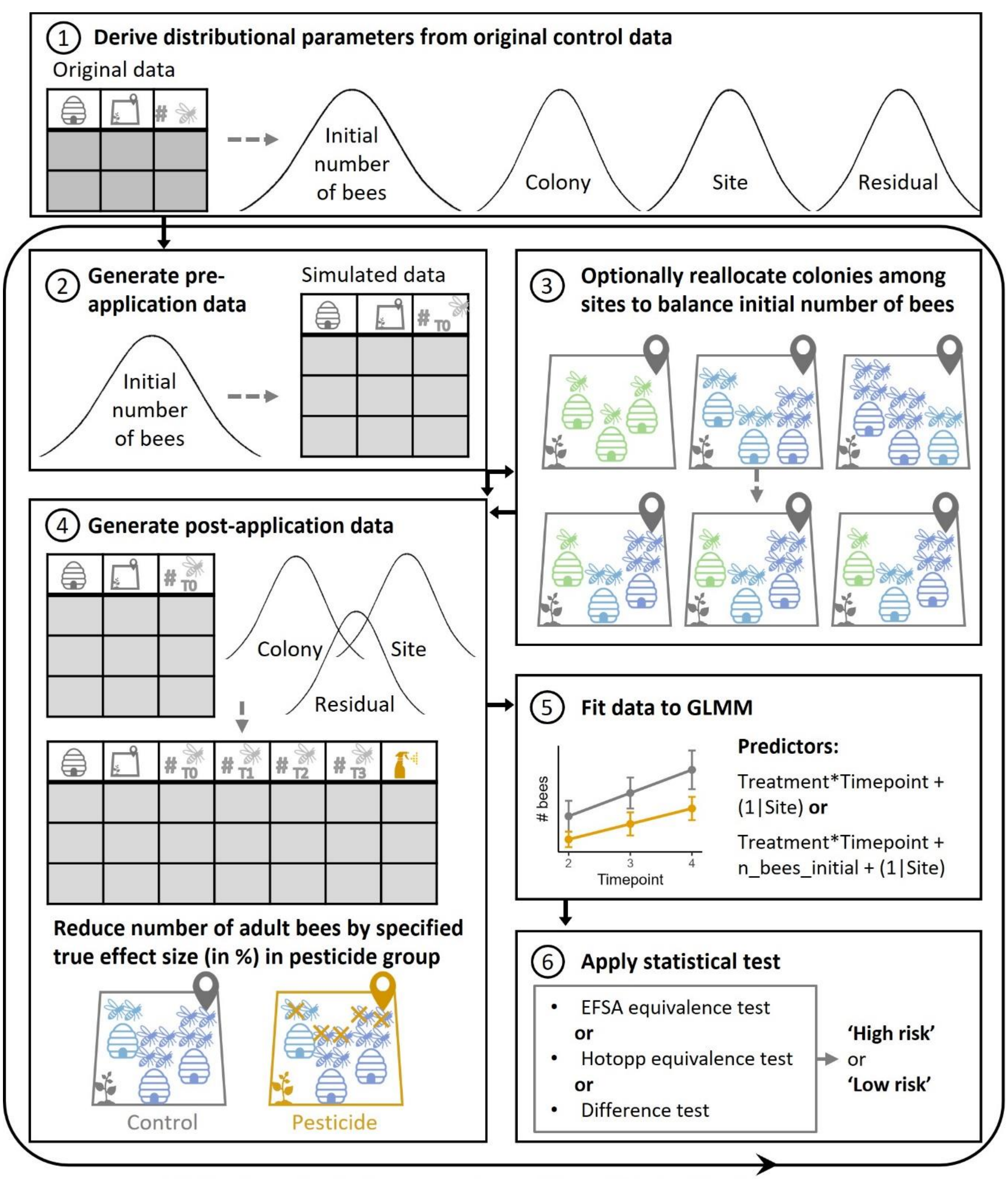

**Fig. 1. Simulation procedure for pesticide effects on honeybee colonies.** (1) Distributional parameters were derived from the original control data, which were used in a risk simulation process. (2) The initial number of bees was simulated for a dataset of a desired number of sites and colonies. (3) Under the design and combined adjustment strategies, colonies were reallocated among sites using anticlustering randomisation to balance the initial number of bees. (4) Data for time points after the pesticide application day were simulated (without treatment effects) through a parametric bootstrap. Treatments (control & pesticide) were randomly allocated either to individual sites or groups of sites formed during the balancing process to minimise differences in initial number of bees. Pesticide effects of predefined size were applied to all colonies at treatment sites, reducing the number of bees at timepoints 2, 3 and 4 by the same specified percentage. (5) Simulated datasets were analysed with generalised linear mixed models (GLMMs) to estimate treatment effects. (6) Risk classification was based on one-sided tests (EFSA equivalence test, Hotopp modified equivalence test, or difference test). The process was repeated 5,000 times per covariate adjustment strategy and true effect size to calculate the percentage of simulations in which the pesticide was classified as 'low risk'.

## 2.3 Deriving distributional parameters from original control data

To enable the parametric bootstrap that allowed for the generation of larger datasets with similar distributional properties as the original dataset, we first derived distributional parameters from the original control data. Initial number of bees (timepoint 1) was simulated separately from later timepoints (2–4), but it served as input for generating bee numbers at those later timepoints. As initial number of bees followed a normal distribution (Fig. A2), we derived the mean and standard deviation from it. For the other timepoints, the original control data (time points 2-4) were fitted to a GLMM using the R function `glmmTMB()` of the `glmmTMB` package (Brooks et al., 2017). The model used a negative binomial error distribution (`family = "nbinom2"`) and contained number of bees as the response, timepoint and log-transformed initial number of bees as fixed effects, and colony identity nested in site identity as random effects. The model fit was validated using the `DHARMa` package (Hartig, 2024; Fig. A3). From this model, standard deviations for the colony and site effects and residuals were derived. The model was later also used for fixed effects predictions (see 2.6 'Generation of post-application data').

## 2.4 Generation of pre-application data

A dataset of desired size was created with identifiers for colonies and sites. For these colonies, initial numbers of bees were simulated from a normal distribution, using the previously derived mean and standard deviation from the first assessment of the original control data and rounded to the next integer. These data fed into the generation of post-application data.

## 2.5 Reallocation of colonies among sites to balance initial number of bees

For the design and combined adjustments, colonies were reallocated using anticlustering randomisation to balance initial number of bees across treatments and sites. This reallocation was done in two steps using a new function called `experimental_allocation()`: First, colonies were allocated to treatment groups (control or pesticide) and then further distributed among sites within each treatment group by additionally supplying the sites through the function's `group_data` argument to ensure similar means and variances across different experimental units. The function uses the `anticlust` package (Papenberg and Klau, 2021) to form anti-clusters, in this case, treatment groups or sites with high within-group heterogeneity and low between-group heterogeneity in terms of the selected covariate (*i.e*. initial number of bees). In our case, we used `method = “local-maximum”` with 10 iterations (`repetitions = 10`) to maximise within-group variance (`objective = “variance”`).

## 2.6 Generation of post-application data

To generate data for timepoints after the day of pesticide application, we first simulated data without treatments. Fixed-effect predictions for each colony-timepoint combination were obtained from the model fitted to the original control data using the `predict()` function of the `stats` package with `re.form = NA` to exclude random effects. Random effects for site and colony, as well as residual variation, were added by sampling from normal distributions with standard deviations estimated as previously described and mean zero. If balanced allocation had been applied, the site effect was assigned based on the post-allocation site identity.

Treatments (control or pesticide) were assigned to sites such that half of the sites received either treatment. This assignment was made either fully at random or using the balanced assignment derived from anticlustering randomisation, which randomly allocated treatments to anticlusters. Treatment effects of predetermined size were applied to all sites of the pesticide group by reducing the number of bees per colony by a percentage corresponding to the specified effect size. The same true effect size was applied to all timepoints of a simulated study.

The simulated data closely matched the original control data, as confirmed by visual comparison of observed values, mean estimates, and 95% confidence intervals (Fig. A4).

## 2.7 Model fitting

The generated data were fitted to a GLMM with a negative binomial error distribution (`family` = “nbinom2”) with number of bees as response and an interaction between treatment (categories: control or pesticide) and timepoint (categorical) including main effects as fixed effects and colony identity nested in site identity as random factors. Under the statistical and combined adjustment strategies, the model adjusted for initial number of bees by including its log-transformed value as an additional fixed effect.

## 2.8 Model evaluation

Models were evaluated using the EFSA equivalence test, the Hotopp equivalence test and/or a difference test to classify pesticides as ‘low risk’ or ‘high risk’. All three tests were one-sided and conducted using the `emmeans()` function of the `emmeans` package (Lenth, 2025). EFSA recommends analysing the treatment effect across timepoints if there is no significant interaction between treatment and timepoint; otherwise, the treatment effect shall be analysed for each assessment timepoint (EFSA 2023). Typically, the treatment effect has been estimated based on the assessment time point at the end of the exposure period, *i.e*. the crop bloom (EFSA 2013; Rundlöf et al., 2015; Woodcock et al., 2017). Therefore, we analysed both the treatment effect for the final time point of the exposure period (by specifying time point in the `by` argument of `emmeans()` and filtering for the fourth assessment time point) and the marginal treatment effect across time points. Results for the final assessment time point are presented in the main text, and a comparison with analyses across assessment time points is provided in the Supplementary Material; both approaches yielded very similar results (Figures A5–A8). *p*-values were calculated using the `test()` function of the `emmeans` package.

For the EFSA equivalence test, the `null` argument in `emmeans()` was set to log(0.9), reflecting the -10% effect size threshold defined as the specific protection goal, on the scale of the linear predictor (the log scale). With an $\alpha$ level of 0.2 (*i.e*., 20%, as proposed by EFSA and used here unless specified otherwise), this corresponds to a 60% confidence interval (100% - 2 × 20% = 60%) of the log ratio (the estimated log marginal mean of the pesticide group minus that of the control) being larger than log (0.9). The *side* argument was set to “>” to ensure that the lower bound of the confidence interval was tested against the threshold, reflecting the one-sided nature of this equivalence test, which is actually a non-inferiority test of the pesticide treatment compared to the control (see Table 2 for explanations of specifications). A significant equivalence test result ($p < \alpha$) led to the pesticide being classified as ‘low risk’.

For the Hotopp equivalence test, the log ratio representing the treatment effect was calculated as the difference between the estimated log marginal mean of the pesticide group and the lower limit of the 90% confidence interval of the control group's estimated log marginal mean. 'Low risk' was concluded if the lower bound of the 60% confidence interval (*i.e*. when $\alpha = 0.2$) for this adjusted log ratio, calculated using the standard error of the regular log ratio, exceeded log(0.9). This meant there was no overlap between the confidence interval of the effect size and the boundary of log(0.9). For tests using different $\alpha$ levels, the width of the confidence interval for the adjusted log ratio was adjusted accordingly as 100% - 2 × $\alpha$ (in percent).

Following EFSA's recommendation, failure to demonstrate 'low risk' was interpreted as 'high risk' (EFSA 2023). In the difference test, 'high risk' was concluded if the treatment effect was significant ($\alpha$ = 5%; i.e. $p<0.05$).

## 2.9 Comparison of EFSA and Hotopp equivalence tests

To compare the ability of the EFSA and Hotopp equivalence tests to distinguish between truly low-risk and truly high-risk pesticides, we simulated risk assessments across a range of $\alpha$ levels. In EFSA's equivalence test, the $\alpha$ level controls the rate of 'low risk' classifications at the 10% threshold, but not in Hotopp *et al.*'s test. To test whether one test is superior to the other when the level of protection against falsely classifying high-risk pesticides as 'low risk' is aligned, we simulated risk assessments with $\alpha$ levels ranging from 1% to 50% (in 1% increments until 5%, then 5% increments until 50%) for both a pesticide with true effect size of zero (*i.e.* a low-risk pesticide) and a pesticide with a true effect size of 11 percent (*i.e.* a high-risk pesticide). We simulated this assuming 10 sites per treatment and 6 colonies per site (as recommended by EFSA, see Table 1).

We then plotted the rate of falsely classifying a low-risk pesticide as 'high risk' (y-axis) against the rate of falsely classifying a high-risk pesticide as 'low risk' (x-axis). In such a Pareto plot, the test with the curve closer to the bottom-left corner is considered superior (Mair et al., 2020). This comparison allows determining which equivalence test has a better balance between false trusts and false mistrusts and is therefore inherently better at discriminating between truly low-risk and truly high-risk pesticides. In addition, it allows for determining the $\alpha$ level at which the best balance (dependent on the costs assigned to each of the errors) between both error rates is achieved.

To evaluate the ability of the two equivalence tests (EFSA and Hotopp) to identify high-risk pesticides, risk simulations were conducted for effect sizes ranging from -10% to -20% in 2% increments and for numbers of sites per treatment ranging from 4 to 20 in increments of 2. The number of colonies per site was assumed to be 6 in all simulations. No adjustment for initial number of bees was made. The $\alpha$ level was set to 20% as recommended by EFSA (EFSA 2023).

### 2.10 Comparison of adjustment strategies for initial number of bees

To determine how many sites per treatment would be necessary to achieve 80% power, we examined how the probability of classifying a pesticide as 'low risk' is affected by the adjustment strategy for initial number of bees. Using the EFSA equivalence test, we tested this with number of sites per treatment ranging from 4 to 20 in increments of 2 and from 25 to 45 in increments of 5 and the following true effect sizes: 0% and -5%. We chose a true effect size of -5%, as that is a negative but tolerable effect. Number of colonies per site was again fixed at six. The crossing points between the probability of classifying a pesticide as 'low risk' and the 80% threshold were estimated by linearly interpolating between the two nearest evaluated site numbers.

As the revised EFSA guidance is currently a non-binding recommendation which has received criticism for requiring too many sites, we compared it against the former guidance. For this, we repeated the simulations with the same covariate adjustment strategies and sample sizes for a -7% true effect size (the threshold under the former EFSA guidance) using difference testing. To determine the minimum number of sites per treatment needed to reach 80% power, we used those simulations to locate the expected range and re-ran simulations around that point to identify the smallest number of sites per treatment where at least 80% of runs yielded $p<0.05$.

All analyses of this study were done in R version 4.5.0 (R Core Team, 2025), and the R scripts are made available on GitHub (https://github.com/Dimitry-Wintermantel/equivalence-testing-and-balanced-allocation-paper/tree/preprint).

## 3 Results

### 3.1 Comparison of EFSA and Hotopp equivalence tests

First, we compared the Hotopp and EFSA tests in terms of their trade-offs between false trust and false mistrust rates, we varied the $\alpha$ level of the tests, allowing us to determine whether one test is intrinsically better than the other in discriminating between high-risk and low-risk pesticides. We find that no test was Pareto-superior to the other, meaning that if their protection against misclassifying high-risk are aligned by adapting the $\alpha$ levels, the tests are equally good in classifying a harmless pesticide (true effect size = 0%) as 'low risk'. However, setting the $\alpha$ level to approximately 2% in Hotopp *et al.*'s test was required to reach the same protection level as the EFSA equivalence test with $\alpha$ = 20% (Fig. 2A; Fig. A5). For a smaller sample size (4 sites per treatment and 4 colonies per site), $\alpha$ = 3% in Hotopp *et al.*'s test corresponded approximately with $\alpha$ = 20% in the EFSA equivalence test (Fig. A5).

Hence, the Hotopp test is calibrated differently: For a given $\alpha$ level, it produces higher false-trust rates than the EFSA test, but the overall trade-off between false 'low-risk' and false 'high-risk' conclusions remains the same. Essentially, this means that while the Hotopp test nominally has an $\alpha$ level of 0.2, its effective $\alpha$ level as seen from a standard equivalence test is much higher. The conversion rate between nominal $\alpha$ levels is, however, not fixed even in a balanced design with similar variability within the treatments, but weakly depends on sample size.

Next, we compared the EFSA and Hotopp equivalence tests in their ability to correctly identify high-risk pesticides. Across a range of effect sizes exceeding the -10% threshold, the EFSA test consistently showed lower false trust rates than the Hotopp test (Fig. 2B), *i.e.*, fewer high-risk pesticides were falsely classified as 'low risk'. Because the $\alpha$ level, which was set to 20%, reliably controls the false trust rate in the EFSA equivalence test, the probability of a false 'low risk' classification under the EFSA test remained below 20% for all tested effect sizes. In contrast, the modified equivalence test proposed by Hotopp *et al.* yielded considerably higher false trust rates under the same adjustment strategy. For example, in a study with six sites per treatment—typical of current practice—pesticides with a true effect size of -12% were misclassified as 'low risk' in half the cases and pesticides with a true effect size of -15% in a third of cases, if the Hotopp test was used based on the final assessment time point (results for analyses across time points during the exposure period are similar and can be found in Fig. A6). Even for the sample size recommended in the revised EFSA guidance of 10 sites per treatment (and 6 colonies per site), 48% of pesticides with a -12% effect size and 26% of pesticides with a -15% effect were misclassified as low risk using the Hotopp test. In fact, for studies with 20 sites per treatment, still more than 40% of tests with pesticides causing honeybee colony size reductions of 12% would pass the bar to be considered 'low risk'. At least 15 sites were needed to keep the misclassification rate for pesticides with a true effect size of -15% below 20%.

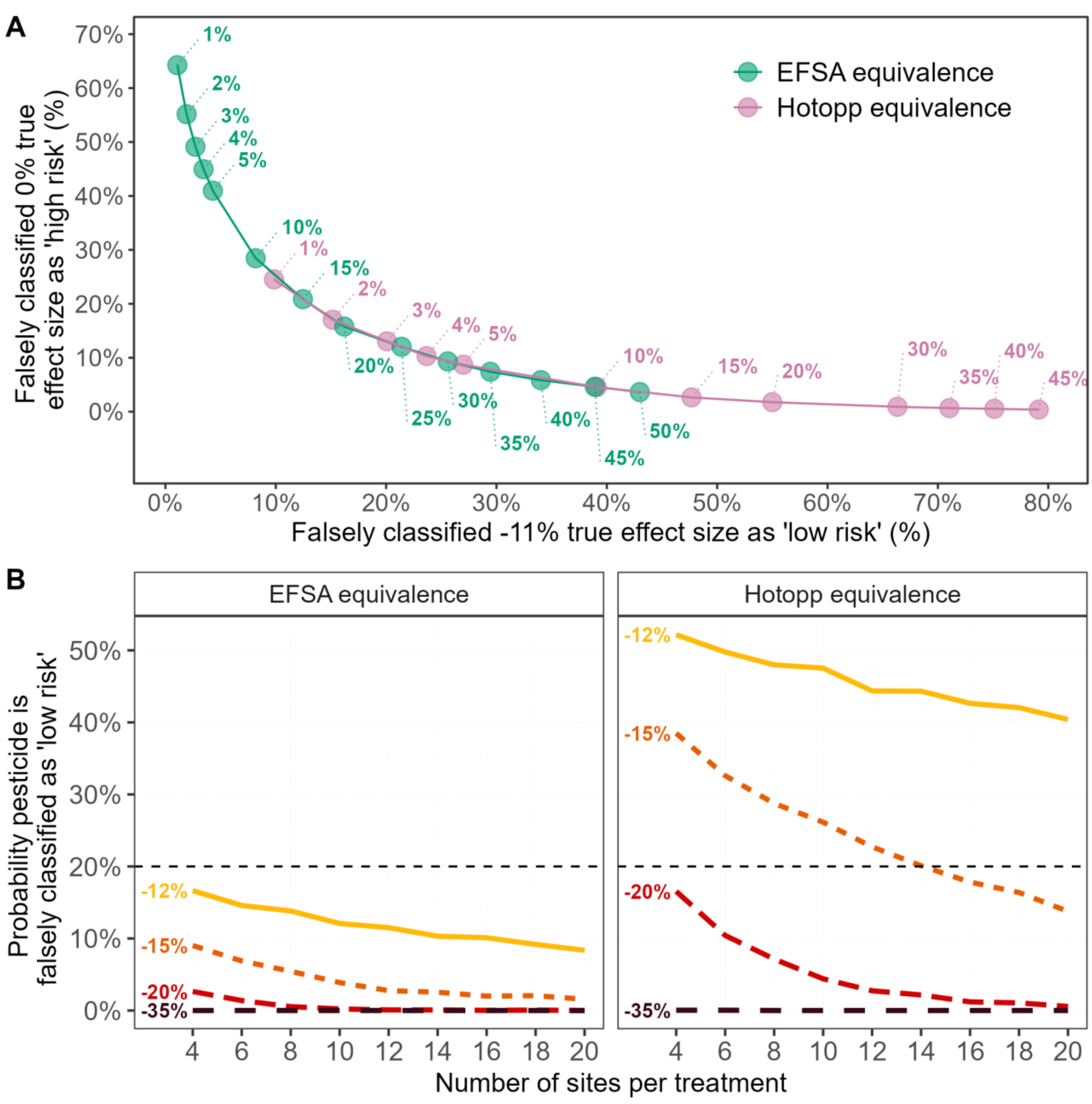

**Fig. 2. Comparison between equivalence tests proposed by EFSA and Hotopp *et al.* in terms of false classification rates. A)** Rates at which pesticides are falsely classified as 'high risk' or 'low risk' when the true effect size is 0% and -11%, respectively. Labels indicate different $\alpha$ levels. **B)** Probability of a pesticide being classified as 'low risk' depending on the number of sites and effect size, assuming six colonies per site**.** In all analyses, the threshold was set to -10%, meaning true effect sizes corresponding to colony size reductions < 10% are considered 'low risk'. Both tests were one-sided. All simulations were conducted with 5'000 iterations and with six colonies per treatment and site. Simulations for panel B were done with 10 sites per treatment. Results are shown for the final assessment timepoint of the exposure period.

## 3.2 Site requirements of EFSA's equivalence test under different adjustment strategies

After showing that the test proposed by Hotopp *et al*. offers no benefits over EFSA's equivalence test and instead simply changes the $\alpha$ level, we evaluated the relation between power and number of sites per treatment for low-risk pesticides with true effect sizes of 0% and -5% using EFSA's equivalence test. We also compared power among adjustment strategies for initial number of bees.

Adjusting for the initial number of bees substantially reduced the number of required sites per treatment to reach 80% power using EFSA's equivalence test, irrespective of the adjustment strategy (Fig. 3, Fig. A7). For a pesticide with a true effect size of 0%, 6–7 sites per treatment were needed when initial colony size was adjusted for, compared to 9 sites when it was not, representing a reduction of 2–3 sites. For a pesticide with a true effect size of -5%, the required number of sites dropped from approximately 34 (no adjustment) to just 24–25 when adjusting for initial bee numbers—a reduction of nearly 10 sites per treatment. This indicates that experiments with a typical sample size of current field studies would have sufficient power under the new EFSA guideline when pesticides do not negatively affect honeybee colonies. To reliably show that pesticides with a true effect size of -5% are low risk, the number of fields would have to increase substantially compared to the current practice. However, under the former EFSA guidance from 2013, even more sites would be required: 41 when not adjusting for initial number of bees and 29 when adjusting for it (Fig. 3, Fig. A8).

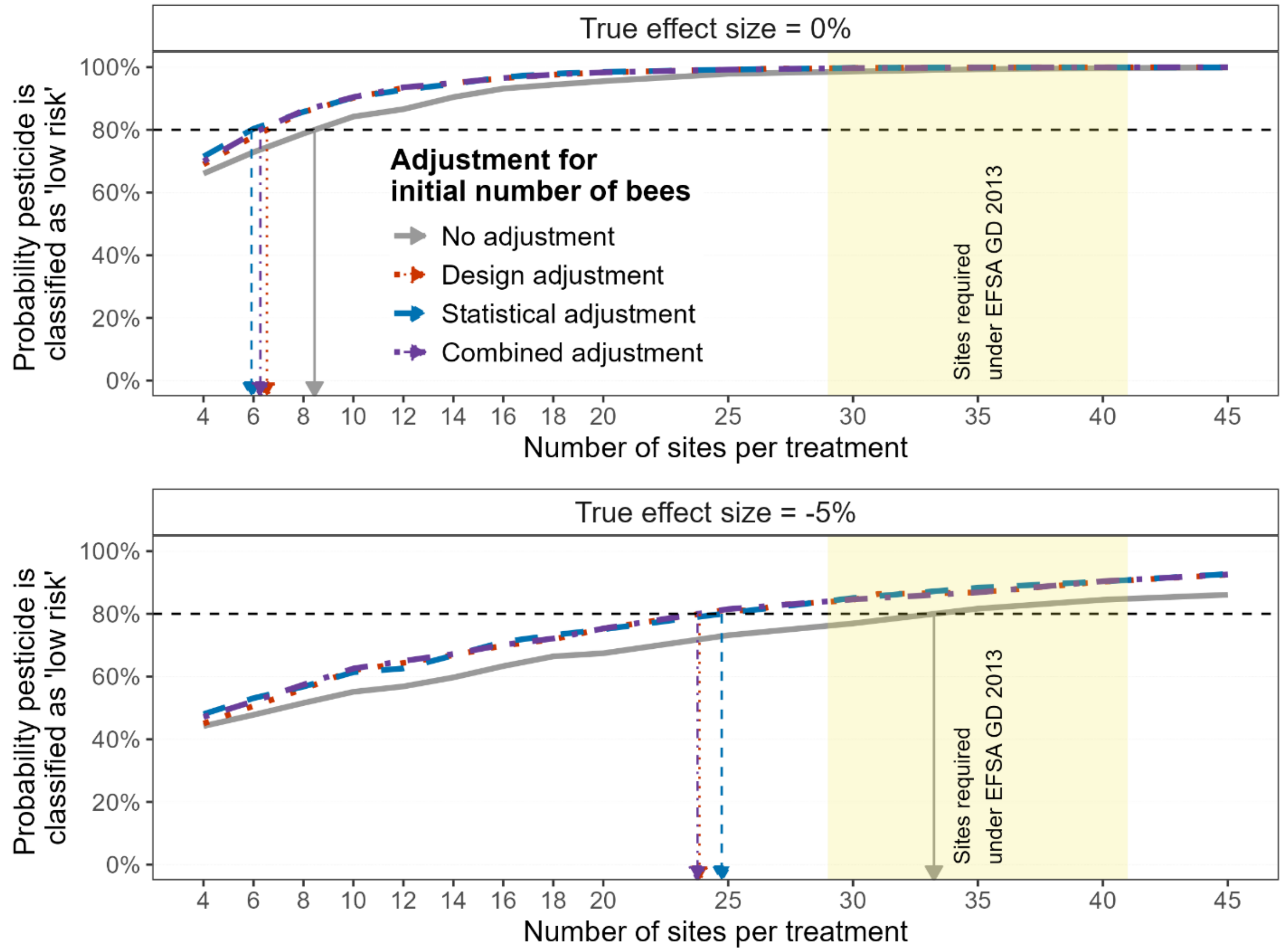

**Fig. 3. Statistical power in relation to number of sites and adjustment strategy.** Curves show the probability that a pesticide is classified as ‘low risk’ (i.e. statistical power) based on EFSA equivalence testing in relation to the number of sites per treatment and adjustment strategy for initial number of bees. Results are shown for two true effect sizes (0% and -5%) and four adjustment strategies: no adjustment, design adjustment (through balanced colony allocation), statistical adjustment (through inclusion of a covariate), and combined adjustment (design and statistical adjustments). The horizontal dashed line indicates the 80% probability threshold. Vertical arrows mark the minimum number of sites needed to reach this threshold for each strategy. Yellow shaded areas indicate the site requirements under the 2013 EFSA guidance document, with the left bound corresponding to the required number when initial bee numbers were adjusted for, and the right bound when they were not. Results are shown for the final assessment time point of the exposure period.

# 4 Discussion

Field studies assessing pesticide risks to honeybees as the primary model species for pollinators often lack sufficient statistical power to protect against effects exceeding the specific protection goal (SPG), which may have contributed to the approval of harmful pesticides in the past. To address this, EFSA recently recommended replacing traditional difference tests with equivalence tests that explicitly evaluate whether effects are below the SPG. While agreeing in principle with the approach, Hotopp *et al.* have argued that this change would make it impractical to demonstrate pesticide safety and have proposed an alternative equivalence test that compares the mean number of bees in treated colonies to the lower confidence bound of the control group mean, aiming to account for natural variability.

Using simulations, we show that the EFSA equivalence test is more conservative, *i.e*. protective against falsely classifying high-risk pesticides as 'low risk', that the site numbers required under EFSA's new guidance were generally lower for pesticides causing a 5% reduction in honeybee colony size or less than those implied by the former guidance, and that adjusting for a covariate could considerably reduce site requirements.

## 4.1 EFSA equivalence testing is better aligned with regulatory goals than Hotopp equivalence and difference testing

Our results demonstrate that the EFSA and Hotopp equivalence tests are essentially identical in their ability to distinguish high-risk from low-risk pesticides. However, the EFSA test is more protective and its nominal $\alpha$ level directly controls the probability that a pesticide is classified as 'low risk' when its effect size equals the -10% threshold and hence ensures that the specific protection goal (SPG) for honeybee colonies is met at the required rate. In contrast, in the Hotopp test, the $\alpha$ level has lost this meaning because the test shifts the baseline for calculating the effect size from the control mean to the lower bound of the 90% confidence interval of the control. For effect sizes defined as the difference in means, the threshold is effectively shifted by half the span of the 90% confidence interval of the control mean. This has three consequences: a) the Hotopp test classifies pesticides with larger true effect sizes as 'low risk', b) reduced precision in estimating only the control mean increases the probability that a pesticide is classified as 'low risk', despite greater underlying uncertainty and c) the nominal $\alpha$ level required to achieve the same level of protection against falsely classifying pesticides with an unacceptably large effect size as 'low risk' vary with sample size. Our results suggest that for the recommended sample size of 10 sites per treatment with 6 colonies per site, a nominal $\alpha$ of 2% in the Hotopp test would be needed to achieve a 20% probability of classifying a pesticide as low risk; meaning the test would pass 10 times more pesticides with a true effect size of -10% than indicated by its nominal $\alpha$.

Hotopp *et al.* (2024) claim the EFSA equivalence does not account for the natural variability of honeybee colonies. However, EFSA based the SPG on the background variability in honeybee colony size as this was the preferred approach among risk managers representing EU Member States (EFSA 2023, EFSA 2021). This means the -10% threshold already reflects expected background variation, not an intent to approve pesticides with true effects close to -10%. Moreover, it is not a flaw but a meaningful feature that higher variability makes it harder to demonstrate equivalence. Imprecise estimates should not lead to regulatory approval. Shifting the thresholds to confidence interval bounds as suggested by Hotopp *et al*. creates incentives to increase background variance (*i.e*., variance in control colonies) to more easily pass equivalence tests, which is clearly problematic. Instead, precision should be improved through increasing the sample size or reducing uncontrolled variation. From a statistical point of view, even pesticides with effect sizes close to the -10% SPG can still be shown to be 'low risk' if the true reduction is below the SPG and the study is sufficiently replicated to achieve the necessary precision, even though this may be unfeasible in practice.

EFSA equivalence testing is also better aligned with regulatory goals than traditional difference testing. Pre-approval pesticide risk assessment is not primarily aimed at demonstrating risks but rather at determining whether a pesticide is safe enough to be approved. By explicitly comparing effect sizes against the SPG, the EFSA equivalence test directly addresses this question. In contrast, difference tests can only reject or fail to reject the null hypothesis of no effect, which is not only dependent on the tested pesticide's effect size but also on the statistical power of the test/study, which in current and past practice has typically been much below the desired levels. The equivalence test ensures that the burden of proof rests firmly with the chemical industry seeking approval of their pesticide and creates an incentive to conduct studies with sufficient power. Higher power reduces type II errors and, consequently, lowers the posterior probability (Colquhoun, 2014) that a pesticide classified as low risk is in fact high risk. Importantly, when using equivalence testing, the probability of classifying a pesticide with a true effect of -35%, a reduction in colony size that is considered critical for colony survival, as 'low risk' would be near zero (EFSA 2013).

## 4.2 Feasibility and sample size implications

Our simulations show that, using the EFSA equivalence test, it is feasible to demonstrate the safety of pesticides with a low true effect size. For pesticides without effect on honeybee colony size (true effect size = 0%), studies of sizes within the range of current practice would suffice: six sites per treatment when adjusting for initial number of bees, or eight sites per treatment without adjustment, in both cases with six colonies per site. This is seemingly in contrast to Hotopp *et al*.' who found in their simulations that a high proportion of harmless pesticides would be classified as 'high risk' using EFSA equivalence testing. However, they simulated studies with lower than typical replication, as they artificially assigned a zero-effect pesticide treatment to half of the control colonies of an existing typical field experiment, effectively halving the sample size compared to a standard design.

For pesticides expected to cause larger effects that do not violate the SPG, such as many insecticides, larger field studies would be required. This scaling of replication with expected risk is consistent with the logic of the tiered-risk assessment scheme and ensures efficient allocation of testing resources. Our simulations indicate that about 24-34 sites per treatment (depending on whether initial number of bees was adjusted for or not) might be required to reliably (i.e. with 80% power) show that a pesticide with a true effect size of -5% is 'low risk'. Nevertheless, the site requirements for such a pesticide (with true effect size = -5%) remain lower than for any pesticide under the former EFSA guideline, which required 80% power to detect a 7% reduction using difference testing and according to our simulations 29-41 sites (depending on whether adjusting for initial number of bees or not).

We show that adjusting for initial number of bees, either through inclusion in the statistical model or by balancing it at the start of experiments across treatments and sites, can considerably reduce the number of sites required to reach acceptable power levels. For example, about ten fewer sites were required to reach acceptable power to show that a pesticide with a true effect size of -5% is low risk, potentially making it feasible to demonstrate its safety. In our simulations, statistical adjustment (inclusion of initial number of bees as a fixed factor) and design adjustment (balanced allocation using anticlustering randomisation) worked equally well.

## 4.3 Limitations of the simulation tests

We conducted simulations mimicking field experiments on honeybees where the variance components were based on the control group of a single study (Rolke et al., 2016). This study is representative of regulatory honeybee field experiments in terms of colony preparation approaches, assessment methods, and replication (EFSA 2021) and was explicitly designed for the regulatory approval process. However, the spatial arrangement of colonies in this study was atypical in two regards, which potentially affected colony variation in contrasting ways. First, the colony locations of either treatment were clustered within a distance of less than 4 km (Rolke et al., 2016). While this should be avoided to prevent non-independence of field sites, it may reduce variation. Second, some colonies were placed directly at oilseed rape fields and others 400 meters away, which may potentially increase variation compared to studies where all colonies are placed next to a focal crop. Overall, the colony variation tended to be lower than in some other studies (EFSA 2021; Hotopp et al., 2024; Rolke et al., 2016) but not to the extent that would outweigh sample size differences (EFSA 2021). In an assessment of seven field studies by EFSA, Rolke *et al*.'s ranked fourth in power, placing it directly at the median (EFSA 2021). Consistent with this, our estimates of the number of required field sites closely matched those reported by EFSA. For example, we estimated that demonstrating with 80% power that a pesticide with a true effect size of -5% would require 34 field sites when not adjusting for covariates, compared with EFSA's estimate of 36 (EFSA 2023). For a 0% effect size, the difference was only one site (9 in our analysis and 10 in EFSA 2023).

Beyond properties of the underlying dataset, our simulations rely on assumptions about how variance scales with study size. We assumed that variance components remain constant as the number of sites increases. However, larger-scale studies span wider regions, cover more diverse landscapes and require more colonies that may be less well standardised, which could lead to greater colony variation. EFSA's analysis of background variability did not indicate such a pattern (EFSA 2021), but it could emerge if a certain study scale is exceeded (Hotopp et al., 2024).

Finally, we assumed that balanced allocation does not change the magnitude of random effects, although in practice it may reduce them, implying that the true benefits of balanced allocation could exceed those we report. Experimental colonies are often selected from a larger pool to standardise colony characteristics. The study report and publication by Rolke *et al.* does not provide information on whether or to what extent this was done, but they had additional colonies available that were used to substitute seven colonies damaged during transport. Our R function for balanced allocation also optimises colony selection as part of the balancing procedure. Nevertheless, in our simulations, we assumed the number of available colonies matched the required sample size and thus ignored any additional gains in power or corresponding reductions in the number of required sites that might arise from choosing optimally from a larger pool.

# 5 Guidance for practitioners

We showed that for one predictive variable (initial number of bees), design adjustment (balancing covariate across treatments and sites before the experiment) and statistical adjustment (*i.e.* covariate inclusion as model term) increased power and consequently decreased site requirements equally well. However, compared to statistical adjustment, design adjustment in general has the advantage of not reducing degrees of freedom and being immune to p-hacking because it is implemented before the experiment begins. Design adjustment should, therefore, be favoured over statistical adjustment wherever possible. Statistical adjustment, on the other hand, can in principle adjust for variables not measurable before the study. However, adjusting for variables that may themselves be affected by the treatment (*e.g*. pathogen levels in honeybee colonies measured after pesticide application) can mask indirect treatment effects, because part of the treatment effect may be mediated through these variables and is therefore removed by the adjustment (Laubach et al., 2021). Such adjustments of mediators should generally be avoided in regulatory studies to ensure the total treatment effect is estimated. In addition, statistical adjustment opens the door to selective analysis choices. To minimise this risk, we recommend replacing current pre-notification requirements for regulatory pesticide studies (Regulation (EU) 2019/1381) with stricter pre-registration that includes detailed analysis plans. Making both pre-registered study plans and final study reports publicly available would improve transparency and allow for independent verification of study design and analysis decisions.

To facilitate the implementation of balanced allocation of test subjects using anticlustering, as a form of design adjustment, we introduce an R function (`experimental_allocation()`, Table 2) that can allocate either all available test subjects (*e.g*. bee colonies) or select an optimally balanced subset. It can simultaneously balance multiple covariates that are available before treatment allocation, such as numbers of brood cells, bees, and/or varroa mites for honeybee colonies, and field size, flower density, and/or amount of previously applied pesticides for field sites.

For regulatory studies, we recommend combining one-sided equivalence testing as proposed by EFSA (*i.e.* non-inferiority testing) with complementary testing against the same SPG threshold from the opposite side. This approach allows concluding '(demonstrated) low risk' when effects are shown to be below the SPG and '(demonstrated) high risk' when impacts are shown to exceed it; if neither of the two are demonstrated, then 'high risk not excluded' should in, our view, be concluded (Fig. 4). Regulatory approval should only be granted if a significant equivalence test has demonstrated 'low risk'. Nonetheless, distinguishing between demonstrated risk and failure to demonstrate safety may also be relevant in regulatory contexts, for example, when deciding whether a pesticide should be rejected outright or whether additional evidence should be requested. Both EFSA and Hotopp *et al*. did not make this distinction and interpreted non-significant results from equivalence tests as 'high risk'. However, an underpowered study failing to reach a significant equivalence test result does not demonstrate that a pesticide is high risk, just as a non-significant difference test does not demonstrate the absence of an effect. From a statistical point of view, a non-significant equivalence test result should be interpreted as 'low risk not demonstrated' or more loosely 'high risk not excluded'. For EFSA, concluding 'high risk' from a non-significant equivalence test may be an understandable simplification in the regulatory context: decision-makers face a binary choice (approval or rejection), the pesticides being tested have already shown potential risk in lower-tier assessments, and the applicant has failed to demonstrate safety despite its own interests in doing so. Outside this tiered regulatory framework, however, it is not acceptable and could lead to dangerous misunderstandings if we label a pesticide as 'high risk' solely based on a non-significant equivalence test. Therefore, independent studies assessing pesticide impacts on non-target species (or other treatment impacts) should, in addition to equivalence testing, use traditional difference testing (against the null hypothesis of no effect) to verify that there is evidence for an adverse effect and ideally also testing against the SPG from the opposite side as the one-sided equivalence test to verify effects are large enough to demonstrate 'high risk' in a regulatory sense (Fig. 4).

Various R packages can implement these tests. In Table 2, we provide practical guidance for equivalence testing using the `emmeans` package, specifically for lower-is-worse endpoints like number of bees. For higher-is-worse endpoints (*e.g.* number of dead bees), the interpretation is reversed. For variables where both unusually high and unusually low values are problematic (potentially hormone levels), 'low risk' is verified via a significant two-sided equivalence test, whereas 'high risk' is demonstrated using one-sided testing against the lower and higher SPG bounds.

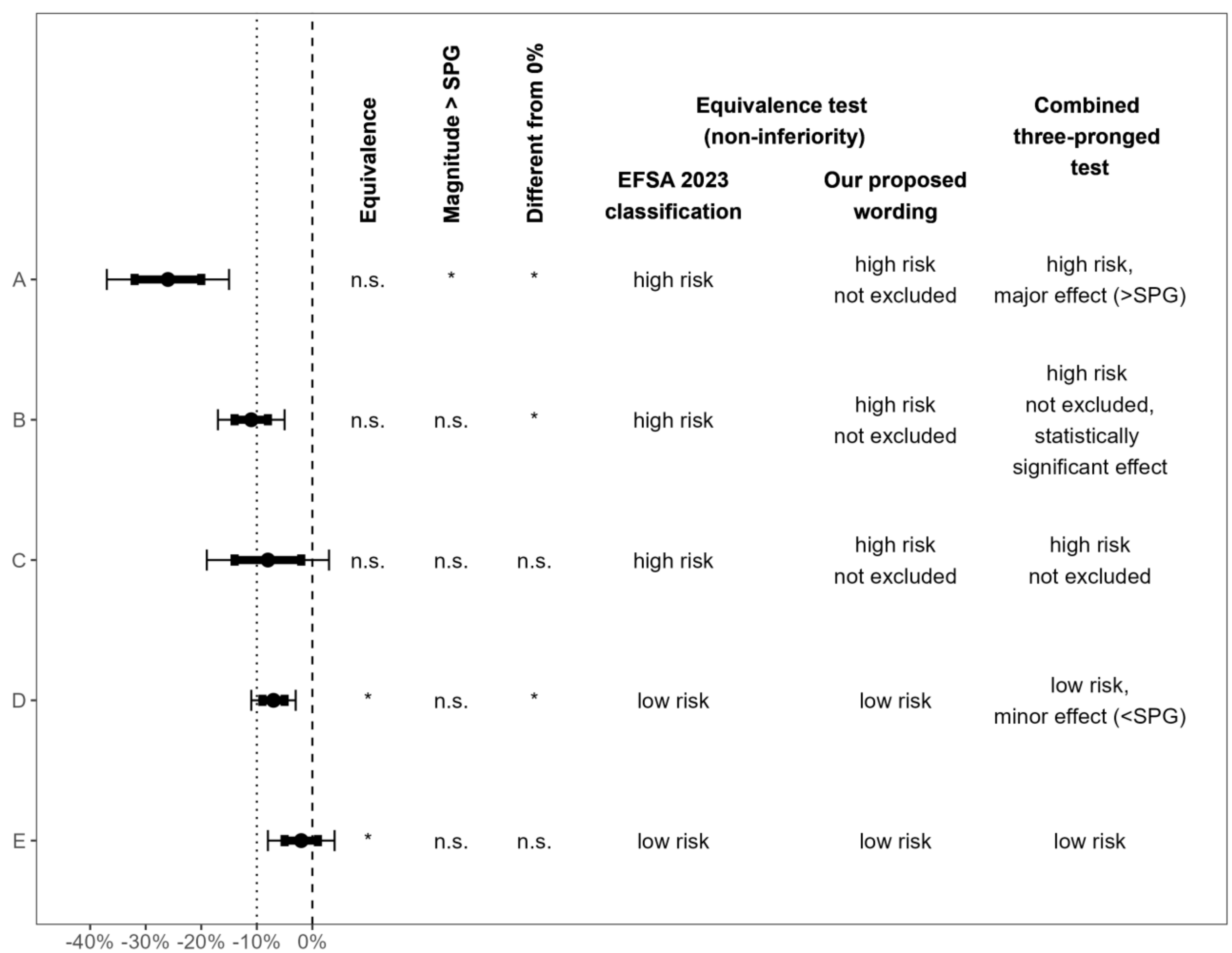

**Fig. 4. Interpretation of pesticide effects based on different statistical tests.** Each row represents a hypothetical pesticide (A–E), with estimated effect sizes shown as points, 90% confidence intervals as thin lines, and 60% confidence intervals as thick lines. A vertical dashed line indicates no effect (0%), and a dotted line indicates the -10% threshold (i.e. the specific protection goal). Significant results of the EFSA equivalence test, test against the SPG from the opposite side, and difference test are indicated with a star (*) and non-significant ones with 'n.s.'. Results are considered significant if $p<0.2$ in the two tests against the SPG and $p<0.05$ in the difference test. The two columns on the right show EFSA's and a strictly statistical interpretation of the equivalence test. The combined test integrates all three test results.

**Table 2. Practical guidance for implementing balanced allocation and equivalence tests against a protection goal in R**

| Objective | How to implement in R? |
|---|---|
| **Balanced allocation** of test subjects (e.g. bee colonies) or experimental units (e.g. sites) to treatments | Apply `experimental_allocation()` to a dataset containing IDs of subjects or experimental units, as well as covariates that shall be balanced across treatments.<br><br>Specify treatment group labels (argument: `treatments`, e.g. c("Control", "Pesticide")), the name of the treatment variable to be created (`treatment_var`) and numeric variables to balance (`covariates`). Optionally, specify factors (e.g. farming system) to balance within (`match_within`) and the number of replicates per treatment (`n_replicates`). The function selects automatically an optimal subset if more subsets are available than needed. See Wintermantel, 2026 for the R function. |
| **Balanced allocation in nested designs** of test subjects to treatments <u>and</u> experimental units (e.g. sites) in nested designs | **Single subject per experimental unit (e.g. one colony per enclosure in semi-field/microcosm experiment):**<br>Step 1 – Use `experimental_allocation()` to assign subjects to treatments.<br>Step 2 – Use `experimental_allocation()` to assign experimental units to treatments (if experimental unit-level covariates matter).<br>Step 3 – Randomly pair subjects with experimental units that have the same assigned treatment.<br><br>**Multiple subjects per experimental unit (e.g. multiple colonies per site in field experiment):**<br>Step 1 and 2 as above.<br>Step 3 – Use `experimental_allocation()` while specifying data sets on subjects (`data`) and experimental units (`group_data`), as well as the treatment variable (`treatment_var`), the experimental units variable (`group_var`) and subject-level covariates (`covariates`; the same as in step 1). |
| **Test against a protection goal** (to verify 'low risk' or 'high risk') | Fit appropriate model, then use `emmeans::emmeans()` to extract the treament contrasts (e.g. `emmeans(model, revpairwise ~ Treatment)$contrasts`).<br><br>Use `emmeans::test()` to test these contrasts against a protection threshold using the `null` argument in one-sided tests (and the `delta` argument in two-sided tests). For models on a multiplicative scale, i.e. with a log link or a log-transformed response, thresholds expressed as ratios (e.g., 0.9 for -10%) must be log-transformed before being passed to `null` (e.g. log(0.9)). For models on an additive scale (identity scale and original units), the threshold should be expressed as the difference in original units (e.g. -100 bees if control mean is 1000 bees for a -10% protection threshold). Specify the significance level ($\alpha$, e.g. 0.2 for EFSA 2023) and the direction of the test:<br>• `side = ">"` (or `"noninferiority"`) to verify 'low risk' for lower-is worse endpoints,<br>• `side = "<"` (or `"nonsuperiority"`) to verify 'high risk' for lower-is-worse endpoints. See `test_equivalence`() within the R script 'Functions to simulate risk assessments.R' provided in the GitHub folder for this study. |

# 6 Conclusions

Our simulations demonstrate that standard equivalence testing as proposed by EFSA is both statistically robust and feasible to implement in honeybee field studies with realistic sample sizes for pesticides with small true effect sizes. By maintaining a clear link to the specific protection goal and controlling the false-trust rate at the defined $\alpha$ level, it ensures transparent, interpretable, and regulation-aligned decisions. Because a 'low-risk' conclusion can only be reached with sufficient statistical power, equivalence testing also creates a direct incentive for applicants to design adequately replicated studies and reduce variance in the test system. Modifying the way the effect size is calculated, as proposed by Hotopp *et al*., is unnecessary and undermines the test's purpose and error rates. Instead, covariate adjustment *e.g*. through balanced allocation of sites to treatments and colonies to sites and treatments, can increase power, reducing site requirements without compromising protection goals. Together, these findings support the EFSA equivalence test as a scientifically sound and practical cornerstone for higher-tier honeybee risk assessments.

These principles generalise beyond honeybees. Any higher-tier ecotoxicity study comparing a treated group with a control may benefit from (i) setting a clear protection goal expressed as an effect threshold, (ii) testing non-inferiority/equivalence against that threshold so $\alpha$ controls false 'low-risk' conclusions, (iii) testing against the same threshold from the opposite side to demonstrate 'high risk' when warranted and (iv) variance-reduction strategies enacted before analysis (e.g. balanced allocation anticlustering randomisation) or specified before analysis (e.g. covariate inclusion).

Broad adoption of these practices would lower the probability of authorising harmful pesticides, aligning regulatory outcomes with environmental protection goals.

## Data availability

The original raw data belong to the Bayer AG, holding the rights to their distribution. However, the study report containing these data can be requested through their transparency initiative (https://www.bayer.com/en/agriculture/transparency-crop-science), as was done for this study. R scripts are made available via GitHub: https://github.com/Dimitry-Wintermantel/equivalence-testing-and-balanced-allocation-paper/tree/preprint

## Declaration of generative AI and AI-assisted technologies in the writing process

During the preparation of this work, the authors used ChatGPT (OpenAI, GPT-4 and GPT-5) to assist with language revision and structuring of text. After using this tool, the authors reviewed and edited the content as needed. All authors take full responsibility for the content of the published article.

## Declaration of competing interests

All authors have nothing to declare.

## Acknowledgements

We thank Anne Bjorkman for her feedback on the conception of this study. This work was supported by the European Union’s Horizon Europe research and innovation framework programme under grant agreement No. 101135238 (WildPosh) and grant agreement No. 101082102 (RestPoll). Additionally, JO was supported by the Svenska Forkingsrådet Formas (2024-00298) and the Strategic Research Area “Biodiversity and Ecosystem Services in a Changing Climate” BECC funded by the Swedish government.

## CRediT authorship contribution statement

D.W.: Conceptualisation, Methodology, Data curation, Formal analysis, Investigation, Software, Visualisation, Writing – original draft. J.O.: Conceptualisation, Investigation, Writing – review & editing. M.M.M.: Conceptualisation, Writing – review & editing. F.H.: Conceptualisation, Methodology, Writing – review & editing.

# Appendix A: Supplementary Material to ‘Equivalence testing in pesticide risk assessment - Evaluation and practical guidance for design, analysis and interpretation’

Dimitry Wintermantel, Julia Osterman, Magdalena M. Mair, Florian Hartig

### Parallel computing

To reduce computational time, we parallelised simulation runs, using the `foreach` package (Microsoft and Weston, 2009), applying the `foreach()` function with the `%dopar%` operator, in combination with the `doParallel` package (Microsoft Corporation and Weston, 2011) for managing cores: the `registerDoParallel()` function registered the parallel backend created with `makeCluster()` for use with `foreach()`. To ensure reproducibility despite parallel execution, we used the `registerDoRNG()` function from the `doRNG` package, which registered a seed number for each simulation run.

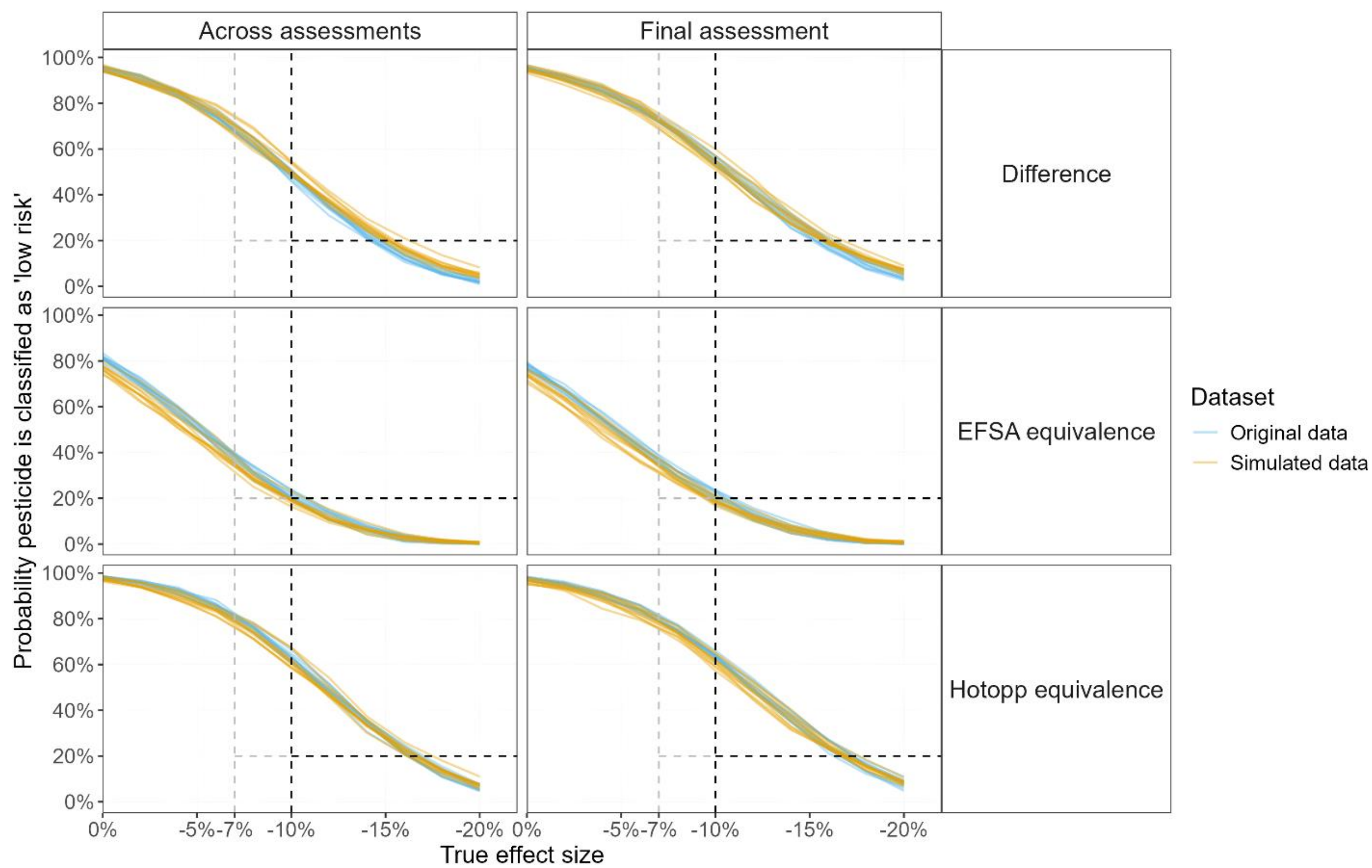

**Fig. A1. Probability that a pesticide is classified as 'low risk' across different tests and true effect sizes, comparing original and simulated datasets.** Each of the 20 curves per panel is calculated based on 500 simulations. Vertical dashed lines indicate the -7% threshold used in the EFSA 2013 guidance (grey) and the -10% threshold used in the EFSA 2023 guidance (black).

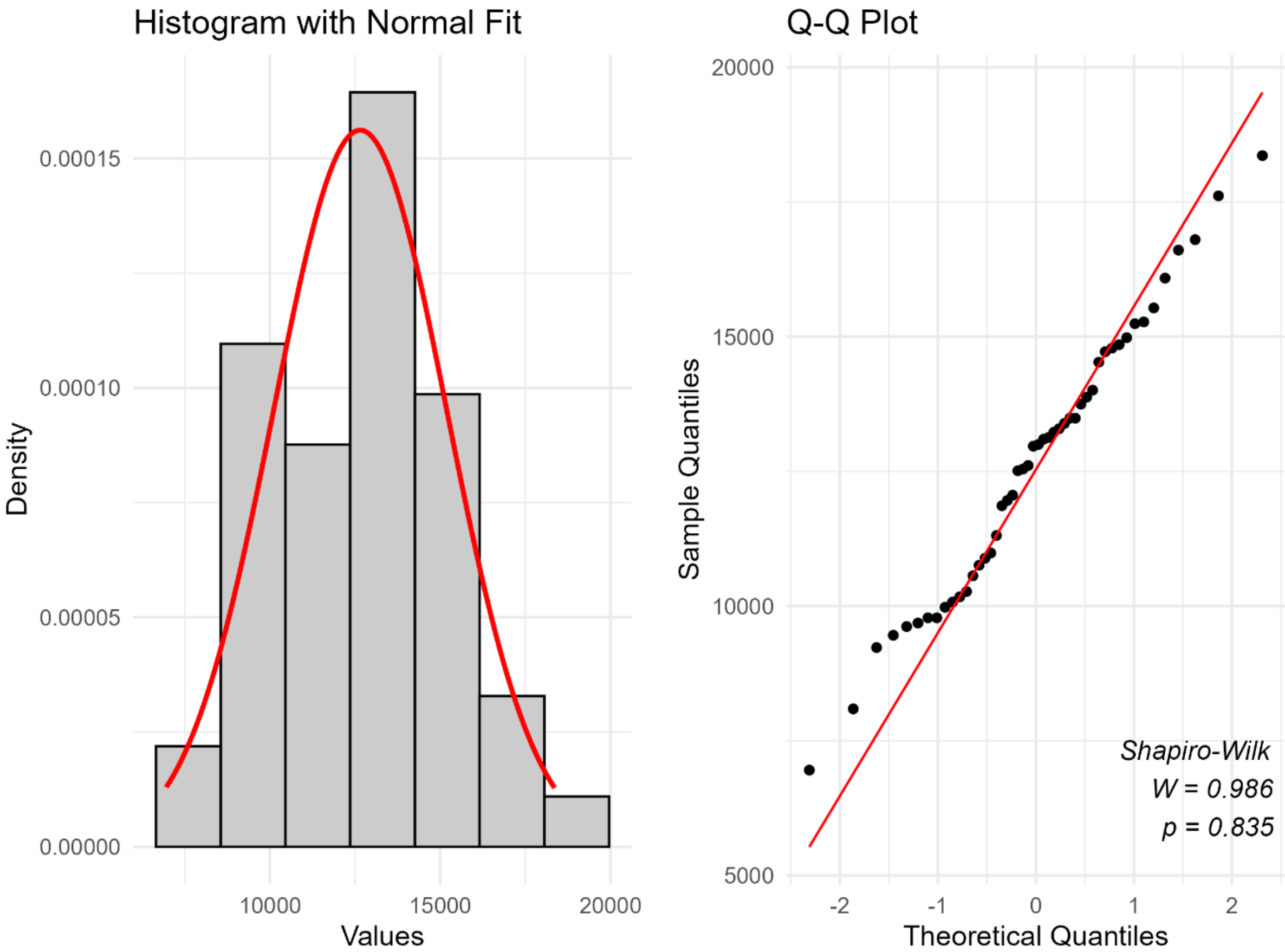

**Fig. A2. Normality assessment of initial number of bees.** The histogram (left) shows the distribution of number of bees at the first assessment with a fitted normal curve (red). The Q-Q plot (right) compares the sample quantiles to a theoretical normal distribution. A Shapiro-Wilk test indicates no significant deviation from normality.

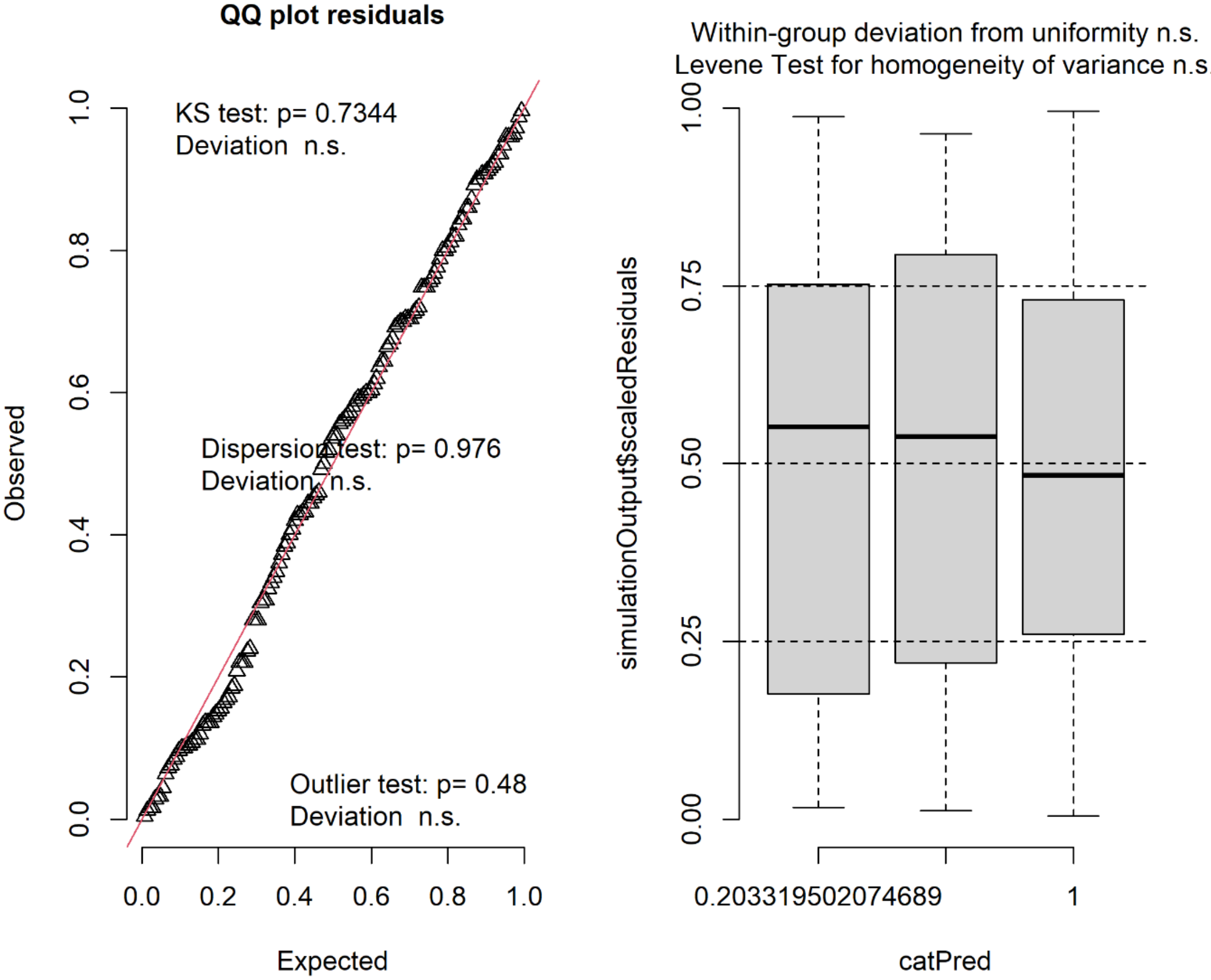

**Fig. A3. Residual diagnostics of the base model.** The plot shows simulated residuals using the DHARMa package for the generalised linear mixed-effects model with number of bees as response assessment as fixed factor and colony identity nested in site identity as random intercepts, including tests for uniformity, dispersion, and outliers to assess the adequacy of model fit.

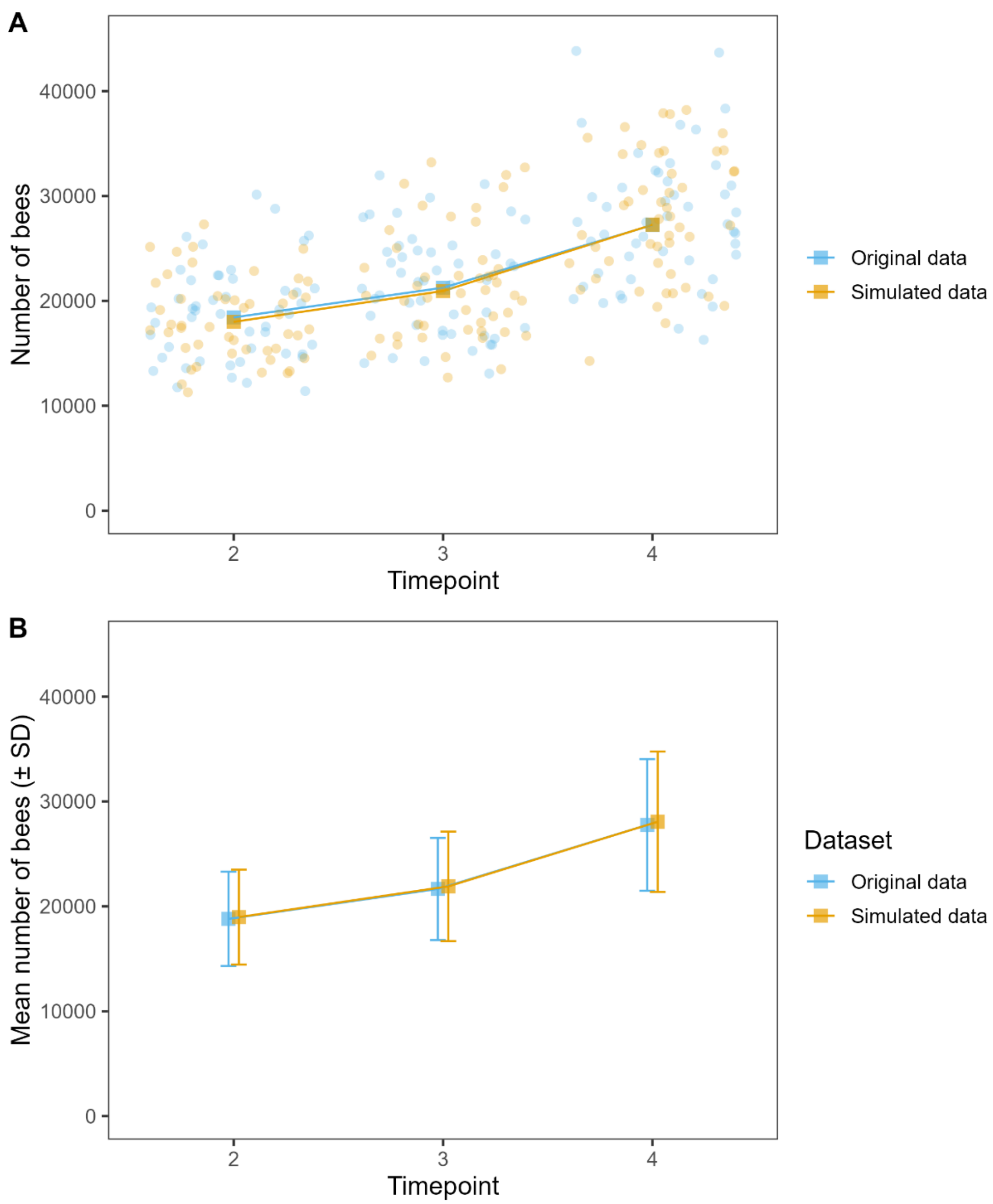

**Fig. A4.** Comparison of original and simulated control datasets. **A**) Estimated marginal means (squares) and observations (circles) for number of bees across post-exposure assessment timepoints from a single simulation iteration compared to the original control data. **B)** Means across 5,000 iterations of estimated marginal means (squares) and corresponding standard deviations (error bars) across post-exposure timepoints compared with the original control data. For both panels, simulated datasets were generated from a GLMM fitted to the original data and matched the original study design in sample size, including the same numbers of sites and colonies per site. No adjustment for initial number of bees was applied.

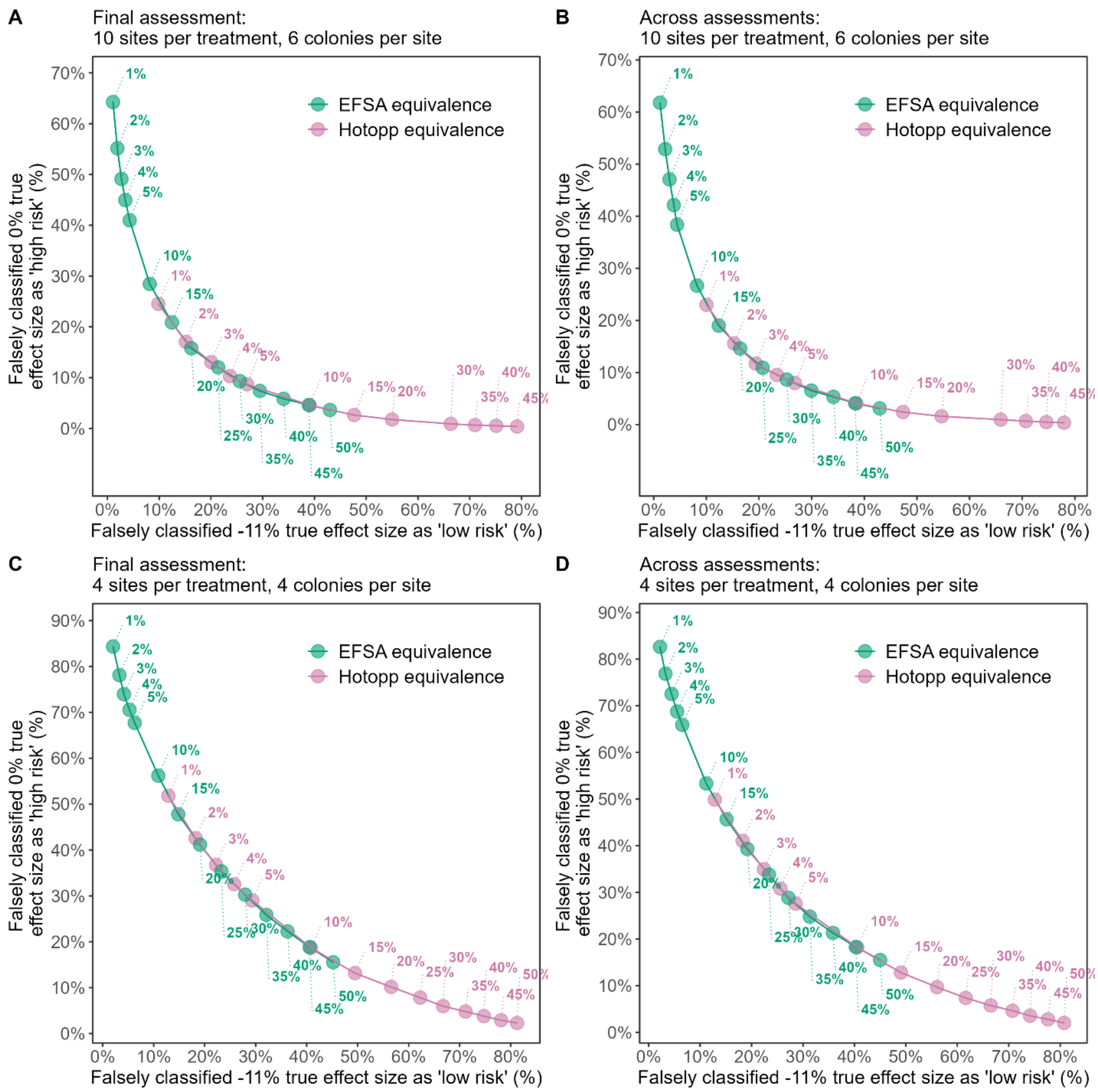

**Fig. A5. Comparison between equivalence tests proposed by EFSA and Hotopp *et al.* in terms of false classification rates.** Rates at which pesticides are falsely classified as 'high risk' or 'low risk' when the true effect size is 0% and -11%, respectively. Labels indicate different $\alpha$ levels. In all analyses, the threshold was set to -10%, meaning true effect sizes corresponding to colony size reductions < 10% are considered 'low risk'. Both tests were one-sided. All simulations were conducted with 5'000 iterations and with six colonies per treatment and site. Simulations in the top row used 10 sites per treatment and 6 colonies per site (**A, B**), whereas simulations in the bottom row used 4 sites per treatment and 4 colonies per site (**C, D**). Results are shown for the final assessment timepoint (A, C) and across timepoints (B, D) of the exposure period.

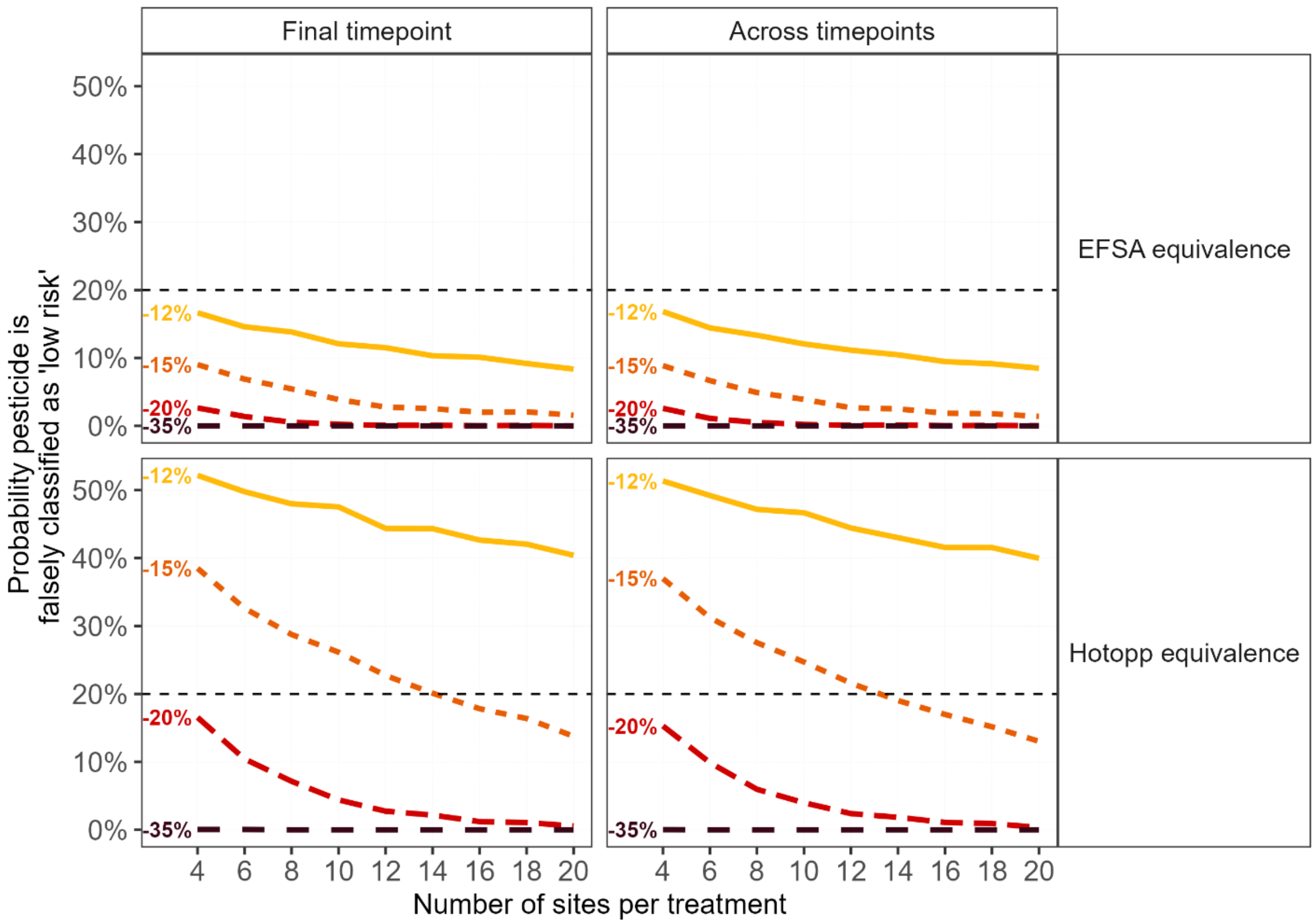

**Fig. A6. Comparison between equivalence tests proposed by EFSA and Hotopp *et al.* in terms of false trust rates.** Probability of a pesticide being classified as 'low risk' depending on the number of sites, effect size and whether the effect was determined only in the final timepoint (i.e. at the end of the oilseed rape bloom) or across timepoints (during oilseed rape bloom). In all analyses, the threshold was set to -10%, meaning true effect sizes corresponding to colony size reductions < 10% are considered 'low risk'. Both tests were one-sided. All simulations were conducted with six colonies per treatment and site.

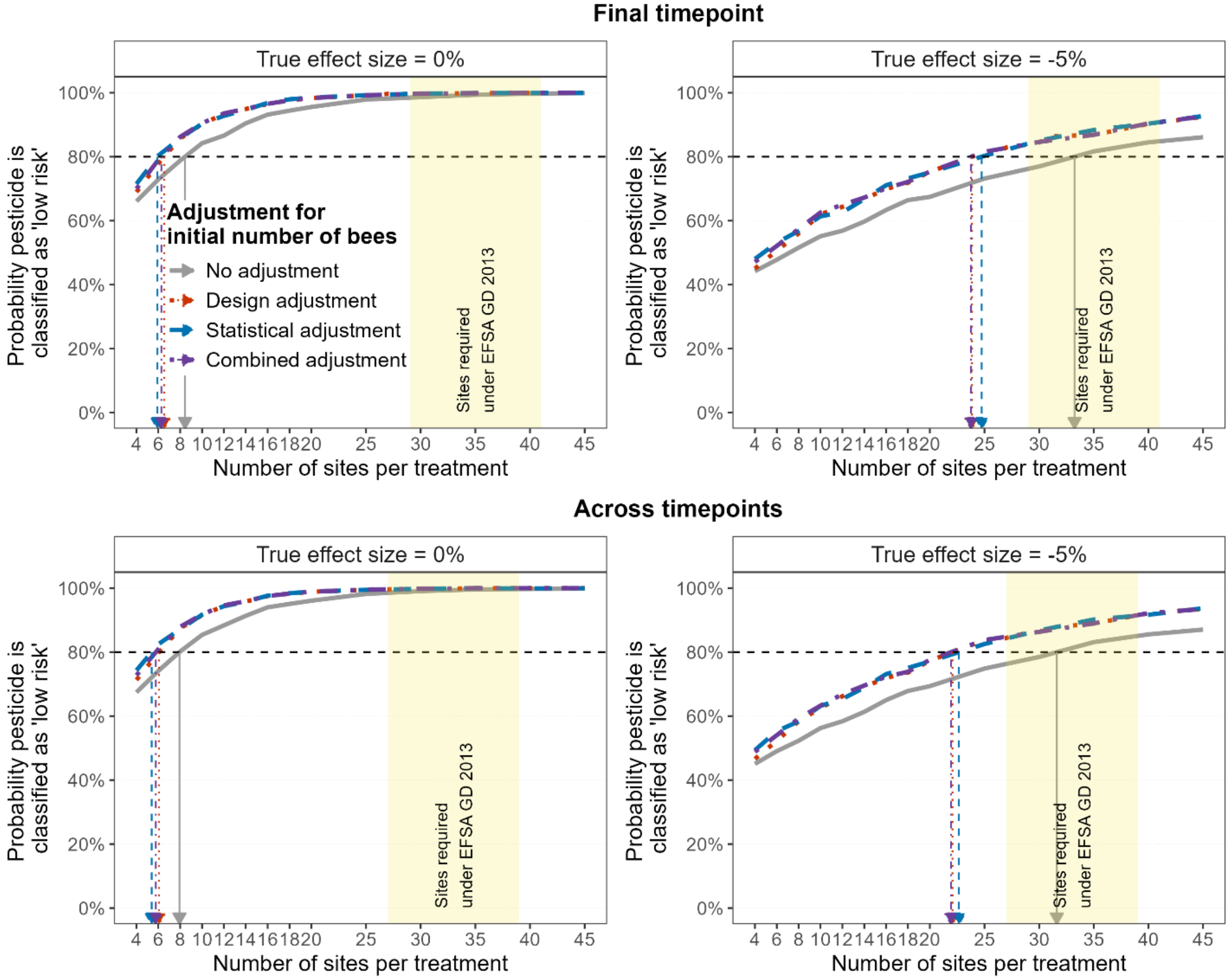

**Fig. A7. Statistical power in relation to number of sites and adjustment strategy.** Curves show the probability that a pesticide is classified as 'low risk' (i.e. statistical power) based on EFSA equivalence testing, shown in relation to the number of sites per treatment and the adjustment strategy for initial number of bees. The top row shows results for the final assessment timepoint of the exposure period, and the bottom row shows results for analyses across assessment timepoints. Results are shown for two true effect sizes (0% and -5%) and four adjustment strategies: no adjustment, design adjustment (through balanced colony allocation), statistical adjustment (through inclusion of a covariate), and combined adjustment (design and statistical adjustment). The horizontal dashed line indicates the 80% probability threshold. Vertical arrows mark the minimum number of sites needed to reach this threshold for each strategy. Yellow shaded areas indicate the site requirements under the 2013 EFSA guidance document, with the left bound corresponding to the required number when initial bee numbers were adjusted for, and the right bound when they were not.

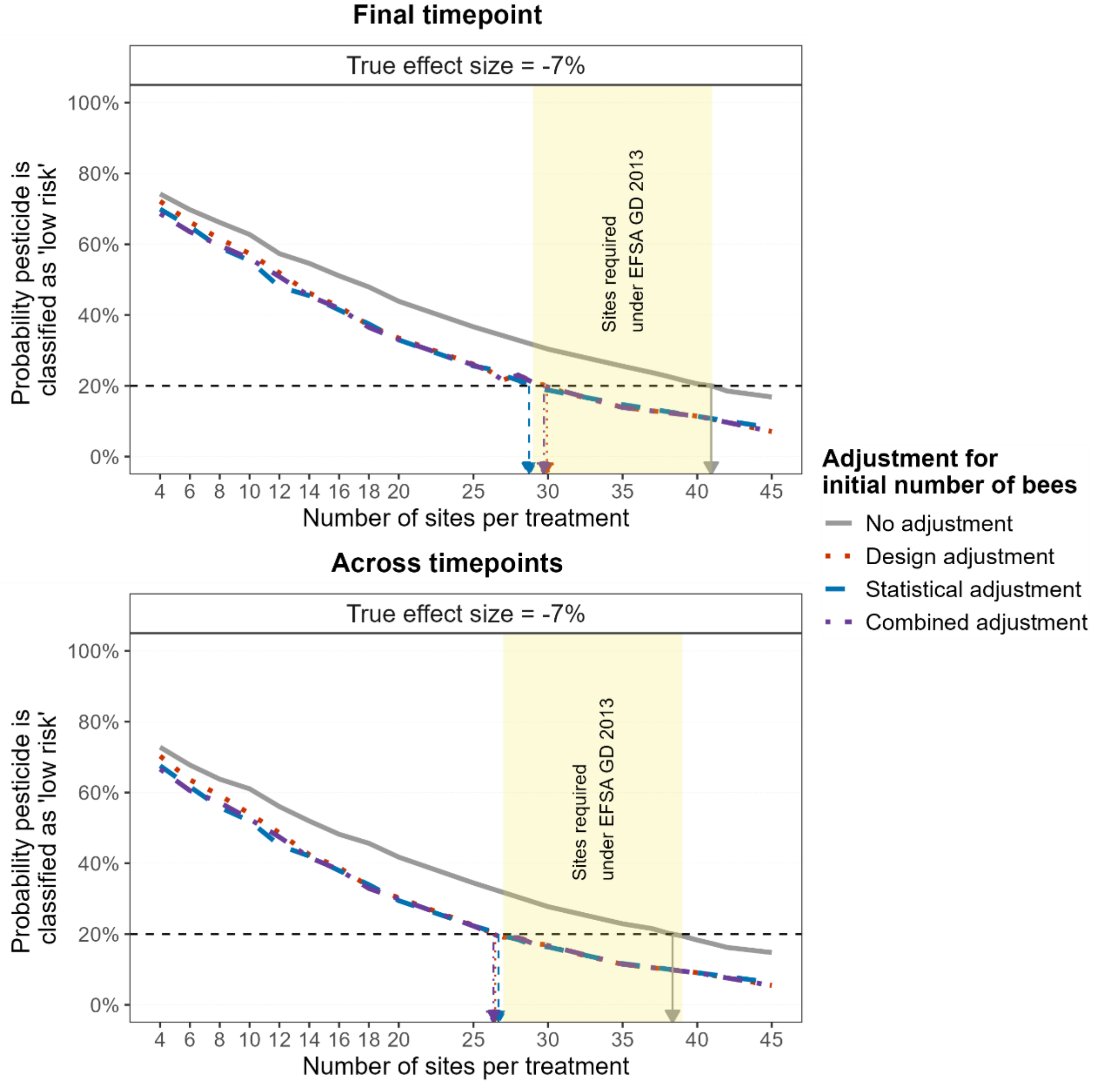

**Fig. A8. Probability of 'low risk' classification in relation to number of sites and adjustment strategy.** Curves show the probability that a pesticide is classified as 'low risk' based on difference testing, shown in relation to the number of sites per treatment and the adjustment strategy for initial number of bees. The top panel shows results for the final assessment timepoint of the exposure period, and the bottom panel shows analyses across assessment timepoints. Results are shown for two true effect sizes (0% and -5%) and four adjustment strategies: no adjustment, design adjustment (through balanced colony allocation), statistical adjustment (through inclusion of a covariate), and combined adjustment (design and statistical adjustments). The horizontal dashed line indicates the 20% probability of 'low risk' classification corresponding to 80% power. Vertical arrows mark the minimum number of sites needed to reach this threshold for each adjustment strategy. Yellow shaded areas indicate the site requirements under the 2013 EFSA guidance document, with the left bound corresponding to the required number when initial bee numbers were adjusted for, and the right bound when they were not.